# A Framework Based Approach for the Development of Web Based Applications

Rachit Mohan Garg, YaminiSood, Balaji Kottana, Pallavi Totlani
Department of Computer Science & Engineering,
Jaypee University of Information Technology, Waknaghat, Solan, H.P, India
{rachit.mohan.garg, eryaminisood, bala1205, pallavi.totlani}@gmail.com

---

Abstract— The sole goal of E-Governance is to allow interaction of government with their citizens in a comfortable & transparent manner. Uniqueness of J2EE makes it a perfect technology for development of any online portal. These involve constancy, easy to replant, construct speedily etc. In this paper we present a procedural approach to develop a web application using the J2EE Struts Framework.

---

## I. Introduction

J2EE is the current field of research as it provides new advancement to the field of web application development framework. Various open source framework which are used for web application development. With the development of technology and communication technology, government wants to modernize their working. E-Governance portal can be used to provide services and important information to citizens through internet, it changes the way of working of government which are operated from traditional approaches to an advanced and more efficient operations. The major advantages of adapting E-Governance / E-Business portal to any government are: removal of location and availability restrictions, reduction of time and money spent, heightening customer services and the provision of competitive advantages [1]-[2]. E-Governance offers a new way forward, helping improve government processes, connect citizens and build interactions with a civil society.

Currently, number of technologies has been developed for constructing E-portal, like ASP.NET and J2EE (Java 2 Enterprise Edition). J2EE is independent of a platform, that's why more and more e-portal adopts this technology. There are many Open Source Frameworks which are based on J2EE technology such as Struts [3], webwork, jsf, echo, Spring [4], real methods, keel, Hibernate [5] and so on. Each framework has its own characteristics, therefore this paper present one web application which is made by Struts Framework [3].

## II. Struts Framework for Web Applications

This paper has taken an example of E-Governance [1] named "Records Collection for India" portal based on above framework. Records Collection for India (RCI) maintains and groups together information from various public services departments that come under the preview of collector's office namely, schools, hospitals etc. Conventionally, the citizen has to go to collector's office in person to get issued certificates like for community, birth, income and driving license etc. This results in wastage of time. Here we provide the online registration facility to apply for various documentation certificates for the citizens so that the public in general has to visit the collector's office only once at the time of submitting relevant documentary proofs or where physical presence is mandatory.

This E-Governance portal has various functionalities. In this section we discuss only login approach which is used by Struts Framework. For this approach, following we require some files namely

- Login,jsp
- Success.jsp
- Faliure.jsp
- Web.xml
- Struts-config.xml
- LoginAction.java
- LoginForm.java
- ApplicationResource.properties

### *A. Web.xml*

For any web application we need to define web.xml file. This file describe first page of that application. This configuration should be done in this as shown fig 1.

```
web.xml
<?xml version="1.0" encoding="UTF-8"?>
<welcome-file-list>
  <welcome-file>welcome.jsp</welcome-file>
</welcome-file-list>
```

Fig. 1 web.xml file.

## B. *Login.jsp*

This page provide user interface for login page. In this page Struts HTML Tags are used for developing user interface. In our application, the login page has one text field to retrieve user name and one password field to get password. And form also contain one submit button, login action class is called when this submit button is clicked. <html: errors /> tag is used to display the error message to user. The login file of our application is shown in fig 2.

```
<div style="color:red">
        <html:errors />
    </div>
<html:form action="/Login"   >
    <p>User Type:<html:select property="choice">
        <html:option value="employee">Employee</html:option>
        <html:option value="citizen" >Citizen</html:option>
        <html:option value="hospital">Hospital</html:option>
        <html:option value="school">School</html:option>
    </html:select></p>
    <p>User Name:<html:text name="LoginForm" property="userName" size="15"/></p>
    <p>Password:<html:password name="LoginForm" property="password" size="15"/></p>
    <p>
<html:submit value="Login"/>
    </p>
  </html:form>
```

Fig. 2 Login.jsp

## C. *Struts-config.xml*

After submitting the form of Login.jsp page, validate method of LoginForm class is called. If there is any error, like username is missing or password is missing, then control is returned back to the input page where errors are displayed to the user.

This complete operation is configured in the action tag of struts-config file. In this web application, “/Login” is action, the input page is “login.jsp” and the corresponding action class is LoginAction.java. The validate method is described in the LoginForm.java file.

## D. *LoginForm.java*

This java file contains validate method; this method is used for checking the entries of username and password of login.jsp page. If there is any error then corresponding error message is displayed to the user. The entire error messages are described in ApplicationResource.properties file. Figure 4 presents the snapshot of the coding written in the LoginForm.java.

```
<struts-config>
    <form-beans>
        <form-bean name="LoginForm" type="com.pawan.LoginForm"/>
    </form-beans>
    <global-exceptions>
    </global-exceptions>
    <global-forwards>
        <forward name="welcome"  path="/Welcome.do"/>
    </global-forwards>
    <action-mappings>
        <action input="/login.jsp" name="LoginForm" path="/Login" scope="session" type="com.pawan.LoginAction">
            <forward name="citizen" path="/citizen_home.jsp" />
            <forward name="employee" path="/employee_home.jsp" />
            <forward name="hospital" path="/hospital_home.jsp" />
            <forward name="admin" path="/admin_home.jsp" />
            <forward name="school" path="/school_home.jsp" />
            <forward name="failure" path="/failure.jsp" />
        </action>
        <action path="/Welcome" forward="/welcomeStruts.jsp"/>
    </action-mappings>
</struts-config>
```

Fig. 3 struts-config.xml.

```
    public ActionErrors validate(ActionMapping mapping, HttpServletRequest request) {
        ActionErrors errors = new ActionErrors();
        if (userName == null || userName.length() < 1) {
            errors.add("userName", new ActionMessage("error.userName.required"));
            // TODO: add 'error.name.required' key to your resources
        }
        if (password == null || password.length() < 1) {
            errors.add("password", new ActionMessage("error.password.required"));
            // TODO: add 'error.name.required' key to your resources
        }
        return errors;
    }
```

Fig. 4 LoginForm.java file

## E. *ApplicationResource.properties*

It contains all error messages which are used in our application. The key “error.userName.required” is used in the validate function to add an error message. By separating error message we can make any change any time without making any changes to the java files or jsp pages.

```
error.userName.required = User Name is required.
error.password.required = Password is required.
```

Fig. 5 ApplicationResource.properties file

## F. *LoginAction.java*

Business logic of web application is written within execute method of LoginAction class. In this file we typecast the ActionForm object to LoginForm, so that we can access the form variables using the getter and setter methods. If the user name and password is correct then we forward the user to the success page else we forward to failure page.

```
package com.pawan;
import java.sql.*;
import java.io.*;
import java.util.*;
import javax.servlet.http.*;
import javax.servlet.http.HttpServletRequest;
import javax.servlet.http.HttpServletResponse;
import org.apache.struts.action.ActionForm;
import org.apache.struts.action.ActionMapping;
import org.apache.struts.action.ActionForward;
public class LoginAction extends org.apache.struts.action.Action
{
    private final static String CITIZEN = "citizen";
    private final static String HOSPITAL = "hospital";
    private final static String SCHOOL = "school";
    private final static String ADMIN = "admin";
    private final static String EMPLOYEE = "employee";
    private final static String FAILURE = "failure";
    public ActionForward execute(ActionMapping mapping,
    ActionForm form, HttpServletRequest request, HttpServletResponse response)throws Exception
    {
        LoginForm loginForm = (LoginForm) form;
        try
        {
            int y=0;
            int ctr,ctr1;
            String uname,choice,password,pass,user;
            uname=loginForm.getUserName();
            password=loginForm.getPassword();
            choice= loginForm.getChoice();
            Class.forName("sun.jdbc.odbc.JdbcOdbcDriver");
            Connection con=DriverManager.getConnection("jdbc:odbc:dsn1");
            Statement st = con.createStatement();
            ResultSet rset;
            HttpSession session = request.getSession(false);
            if(choice.equals("citizen"))
            {
                rset=st.executeQuery("select * from citizen_signup_details");
                while(rset.next())
                {
                    user=rset.getString("emailid");
                    pass=rset.getString("password");

                    ctr=user.compareTo(uname);
                    ctr1=pass.compareTo(password);
                    if((ctr==0)&&(ctr1==0))
                    {
                        session.setAttribute("sessUserName",user);
                        y++;
                    }
                }
                if(y==0)
```

Fig. 6 LoginAction.java file

The procedure described in section 2 provides the implementation of struts framework for the development of E-Governance portals. Here a partial implementation detail of the portal “Records Collection for India” is provided. The overall architecture of the web application is shown in figure 7.

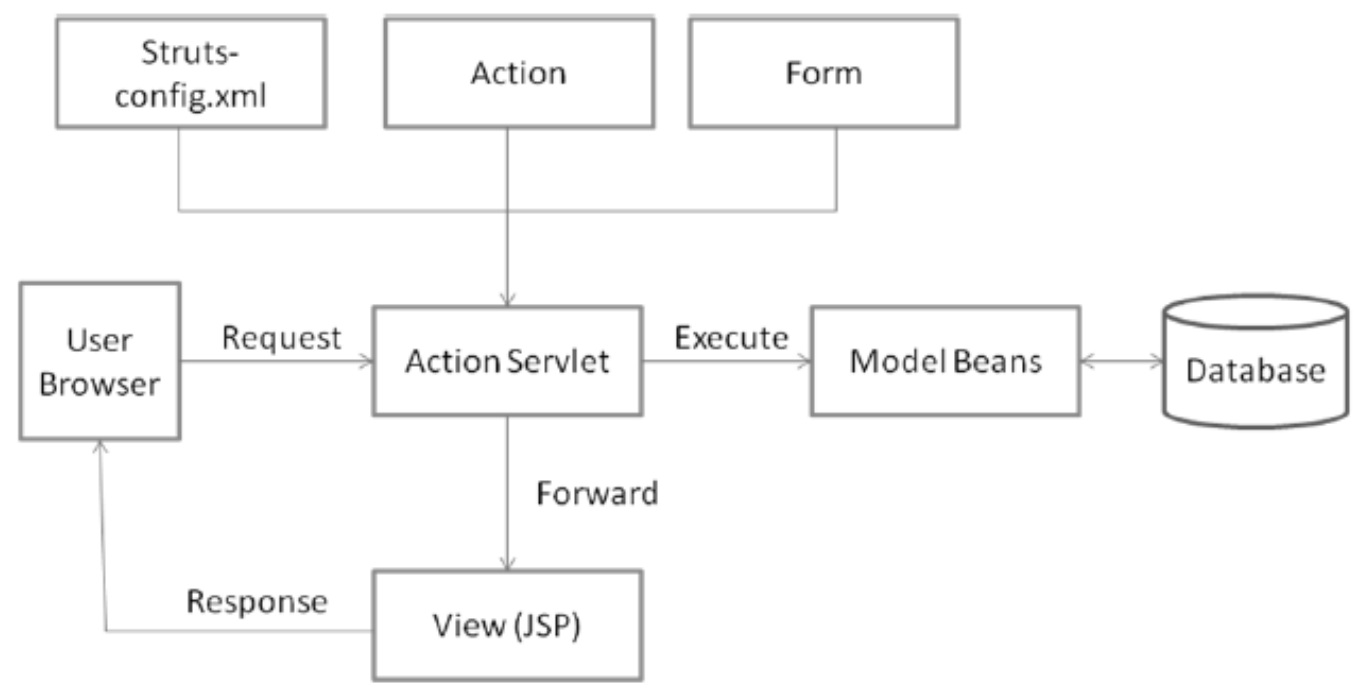

Fig. 7 MVC architecture for the application

The following events happen when the Client browser issues an HTTP request.

- The ActionServlet receives the request.
- The struts-config.xml file contains the details regarding the Actions, ActionForms, ActionMappings and ActionForwards.
- During the startup the ActionServelet reads the struts-config.xml file and creates a database of configuration objects. Later while processing the request the ActionServlet makes decision by refering to this object.

When the ActionServlet receives the request it does the following tasks.

- Bundles all the request values into a JavaBean class which extends Struts ActionForm class.
- Decides which action class to invoke to process the request.
- Validate the data entered by the user.
- The action class processes the request with the help of the model component. The model interacts with the database and process the request.
- After completing the request processing the Action class returns an ActionForward to the controller.
- Based on the ActionForward the controller will invoke the appropriate view.
- The HTTP response is rendered back to the user by the view component.

### *G. Tools Used*

For the implementation of the portal “Records Collection for India” number of tools has been used. Some of the prominent ones are Eclipse Java EE IDE for Web Developers with Struts framework, Ms-Access, Tomcat Server. Eclipse provides the framework for the rapid development of the web application. Ms-Access is used as a backend database for storing the information from the website. The choice of database depended upon its free availability. Tomcat server is used to run the web based application.

## III. CONCLUSIONS

This paper uses "Records Collection for India" as example for implementation using Struts Framework. This application development process is based on J2EE three tier architecture.

In this architecture, the first tier contains the presentation layer i.e. HTML and JSP files, the second tier contains the business logic layer i.e. java files and the third tier contains the data layer i.e. data base.

## IV. FUTURE WORK

Struts framework is a classical implementation of MVC architecture. Hibernate is a powerful technology for persisting data, and it enables Application to access data from any database in a platform-independent manner. Spring is a dependency injection framework that supports IOC. The beauty of spring is that it can integrate well with most of the prevailing popular technologies, thus integrate Struts, Spring and Hibernate is a very perfect pattern.

Future work of this report is to develop an enterprise application which is based on SSH (Struts, Spring and Hibernate).

## ACKNOWLEDGMENT

The authors would like to thank everyone, especially Ms. Shipra Kapoor for always being there to provide a continuous & energious support to them.

## REFERENCES

[1] Hendershot, R. (2007), "E-business Benefits: Learn about the advantages of adopting an eBusiness solution for your small business", www.enetsc.com/EBusinessArticles.htm

[2] E-business and its Advantages (2006), Accessed from: http://onlinebusiness.volusion.com/articles/e-businessadvant ages/

[3] Accessed from: http://struts.apache.org/, Dated Feb 01 2011.

[4] Introduction to the Spring framework, Accessed from: http://www.ibm.com/developerworks/web/library/wa-Spring1/

[5] HIBERNATE - Relational Persistence for Idiomatic Java, Accessed from: http://www.Hibernate.org/hib_docs/ v3/reference/en/html/preface.html

## AUTHORS PROFILE

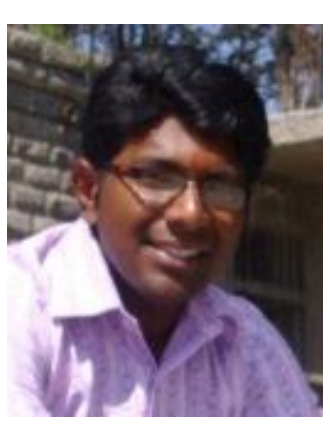

**Rachit Mohan Garg**

The author is pursuing his Post Graduation from Jaypee University of Information Technology with Model Driven Architecture as his research field. He has completed his Engineering from Vishveshwarya Institute of Engineering & Technology; Ghaziabad affiliated to Gautam Buddh Technical University in 2008.

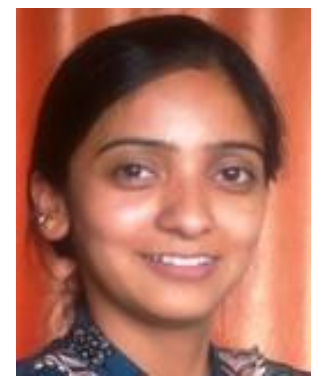

**Yamini Sood**

The author is pursuing her Post Graduation from Jaypee University of Information Technology with Graph Mining as her research field. She has completed her Engineering from Shri Sai College of Engineering & Technology; Badhani, Pathankot affiliated to Punjab Technical University in 2009.

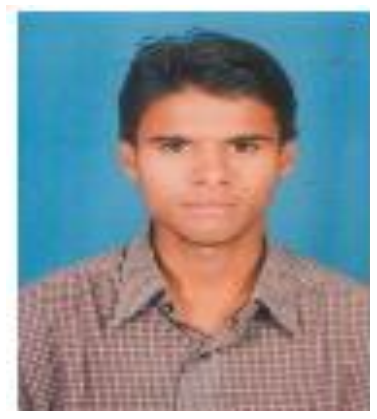

**Balaji Kottana**

The author is pursuing his Post Graduation from Jaypee University of Information Technology. He has completed his Engineering from Maharaj Vijayaram Gajapathi Raj College of Engineering, Chintalavalasa, Vizianagaram in 2007. He is having an industrial experience of 2yrs. in Birlasoft as a software engineer.

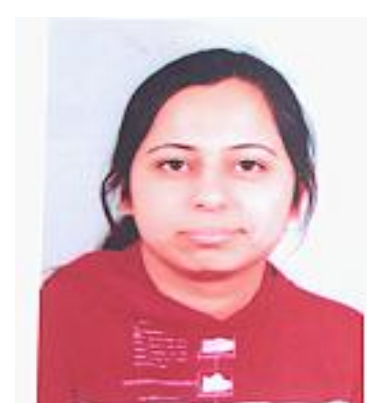

**Pallavi Totlani**

The author is pursuing her Post Graduation from Jaypee University of Information Technology. She has completed her Engineering from Geetanjali Institute of Technical Studies; Udaipur, Rajasthan affiliated to Mohan Lal Sukhadia University in 2008. She also completed her MBA in Finance & IT from Singhania University in 2010.